\documentclass[reprint,amsmath,amssymb,prb]{revtex4-2}
\usepackage{graphicx,subfigure,bm,amssymb,amsmath,hyperref,dcolumn}
\usepackage{color,multirow,supertabular,float}
\usepackage{mathrsfs}
\newcommand{\redq}{\Tilde{q}}
\newcommand{\vk}{\boldsymbol{k}}
\newcommand{\vq}{\boldsymbol{q}}
\newcommand{\redomega}{\Tilde{\omega}}
\begin{abstract}
The spin Coulomb drag effect, arising from the exchange of momentum between electrons of opposite spins, plays a crucial role in the spin transport of interacting electron systems and can be characterized by the exchange-correlation (XC) kernel in the spin channel $K_{\rm XC}^-(q,\omega)$. Using the state-of-the-art Variational Diagrammatic Monte Carlo approach, we compute the Matsubara-frequency-resolved spin XC kernel $K_{\rm XC}^-(q,i\omega_n)$ for the three-dimensional uniform electron gas at sufficiently low temperatures with high precision. In the long-wavelength limit, we identified a singular behavior of the form $A(i\omega_n)/q^2$, confirming the theoretically predicted ultranonlocal behavior associated with spin Coulomb drag. Analysis of this structure in the low frequency region enables precise determination of two crucial parameters characterizing the spin Coulomb drag effect: the spin mass enhancement factor and spin diffusion relaxation time. We observe a significant trend of increasing enhancement of the spin mass factor with decreasing electron density, and provide clear evidence for the suppression of spin diffusion at low temperatures. These quantitative findings advance our understanding of Coulomb interaction effects on spin transport and provide essential parameters for time-dependent density functional theory and spintronics applications.
\end{abstract}
\begin{document}
\author{Zhiyi Li$^{1}$}
	\thanks{These two authors contributed equally to this paper.}
 \author{Pengcheng Hou$^{2,1}$}
	\thanks{These two authors contributed equally to this paper.}
	\author{Youjin Deng$^{1,2,3}$}
        \email{yjdeng@ustc.edu.cn}
	\author{Kun Chen$^{4,5,6}$}
        \email{chenkun@itp.ac.cn}
	\affiliation{$^{1}$ Department of Modern Physics, University of Science and Technology of China, Hefei, Anhui 230026, China}
	\affiliation{$^{2}$ Hefei National Laboratory, University of Science and Technology of China, Hefei 230088, China}
        \affiliation{$^{3}$ Hefei National Research Center for Physical Sciences at the Microscale and School of Physical Sciences, University of Science and Technology of China, Hefei 230026, China}
        \affiliation{$^{4}$CAS Key Laboratory of Theoretical Physics, Institute of Theoretical Physics, Chinese Academy of Sciences, Beijing 100190, China}
        \affiliation{$^{5}$ Center for Computational Quantum Physics, Flatiron Institute, 162 5th Avenue, New York, New York 10010}
        \affiliation{$^{6}$Department of Physics and Astronomy, Rutgers, The State University of New Jersey, Piscataway, NJ 08854-8019 USA}

	\date{\today}
\title{Matsubara-Frequency-Resolved  Spin Exchange-Correlation Kernel for the Three-Dimensional Uniform Electron Gas}
\maketitle
\section{Introduction}
\label{SecI}
Spin transport in interacting electron systems is a fundamental problem with far-reaching implications for spintronics and quantum technologies. The key challenge in describing spin transport lies in the complex interplay between electron-electron interactions and the non-conservation of spin currents. Unlike charge currents, which are conserved, spin currents can be dissipated through the exchange of momentum between electrons of opposite spins. As a result, the up-spin quasiparticle drags along some down-spin electrons in its ``screening cloud".  This mechanism, known as the spin Coulomb drag (SCD) effect \cite{PhysRevB.62.4853,d2010coulomb}, renders the spin dynamics fundamentally distinct from the charge dynamics. 

The SCD effect has been extensively studied across various theoretical and experimental investigations in diverse systems, ranging from solid-state devices to cold atomic gases \cite{SOCSCD,SCDhubbard,weber2005observation,PhysRevB.68.045307,SDoncoldatom,generationofSC,LFCforWeber,NonlocalSCD}. A pioneering experimental confirmation was achieved by Weber et al.  \cite{weber2005observation} in a GaAs quantum well, where measurements of spin trans-resistivity in quasi-two-dimensional electron gases exhibited quantitative agreement with theoretical predictions\cite{PhysRevB.68.045307,LFCforWeber,NonlocalSCD}.

A crucial quantity for characterizing the SCD effect is the exchange-correlation (XC) kernel in the spin channel, defined as the difference between the inverse noninteracting susceptibility and the inverse spin susceptibility of the interacting electrons,
\begin{equation}
    K^-_\mathrm{\rm XC}(\mathbf{q}, \omega) \equiv \chi^{-1}_0(\mathbf{q}, \omega) -\chi^{-1}_S(\mathbf{q}, \omega)
    \label{DefKXC}
\end{equation}
with the superscript `$-$' denoting its antisymmetric nature in the spin channel. 

The XC kernel, parameterized from numerical calculations of the uniform electron gas (UEG) problem, plays a vital role in first-principles calculations of charge and spin properties in real materials. The static and uniform limit of the XC kernel provides a local spin-density approximation (LSDA) for density functional theory (DFT), essential for computing the thermodynamic magnetic properties of real materials~\cite{LSDA,LSDA2}. On the other hand, the frequency-resolved kernel is crucial for the time-dependent density functional theory (TDDFT) to incorporate electron correlations in \emph{ab-initio} predictions of charge and spin dynamics~\cite{TDDFT}. Accurate parameterization of the XC kernel with the correct high-frequency and low-frequency limits is key to developing a TDDFT that captures qualitatively correct spin dynamics~\cite{SCDFT,PhysRevLett.88.056404}.

In the long-wavelength limit, the spin XC kernel exhibits a singular behavior known as ``ultranonlocality"~\cite{PhysRevLett.90.066402,aschebrock2023exact}. In this limit, both the interacting and non-interacting inverse spin susceptibilities exhibit a $1/q^2$-scaling. However, a striking difference emerges between the charge and spin XC kernel. In the charge sector, the conservation of charge currents ensures that the prefactors of this scaling are identical for both the interacting and non-interacting susceptibilities, leading to an exact cancellation in the charge XC kernel. In contrast, the absence of spin current conservation allows Coulomb interactions to renormalize the prefactor of the spin dynamical susceptibility. This renormalization prevents the cancellation that occurs in the charge sector, resulting in a singular behavior of the spin XC kernel,
\begin{equation}
    K_{XC}^-(\mathbf{q}, \omega) = A(\omega)/q^2 + O(1).
    \label{Eq:Low-freq singularity}
\end{equation}

Such singularity was first identified in the high-frequency limit through the third-moment sum rule \cite{PhysRevB.8.200,Liu1991}. Subsequently, the low-frequency limit ($v_F q \ll \omega \ll E_F$ with $v_F$ and $E_F$ the Fermi velocity and the Fermi energy) of the XC kernel was conjectured based on the study of the SCD effect in spin current dynamics~\cite{PhysRevB.62.4853,PhysRevLett.93.106601,XC2dspin,XC3dspin}. The proposed low-frequency expansion is controlled by the spin mass and the relaxation time,
\begin{equation}
    A(\omega) = -\frac{m}{n\tau_{sd}}i\omega -\frac{m_s-m}{m}\frac{m}{n}{\omega^2}+\mathcal{O}(\omega^3),
    \label{eq:XCofsmsd}
\end{equation}
with $n$ the electron density and $m$ the electron mass, $\tau_{sd}$ is the spin relaxation time, and $m_s$ is the spin mass. The quantity $\tau_{sd}$ determines the lifetime of spin currents in the presence of electron-electron interactions and is essential for determining the spin diffusion length and overall efficiency of spin transport, which is proportional to the commonly studied SCD rate in previous work ~\cite{PhysRevB.62.4853,PhysRevB.68.045307,LFCforWeber,NonlocalSCD}, while $m_s$ is a many-body parameter that quantifies the spin current carried by a single quasiparticle \cite{PhysRevLett.93.106601}. Both are two crucial many-body parameters for the SCD effect. 

The ultranonlocality of the spin XC kernel poses significant challenges for accurate parameterization. Traditional methods, particularly the adiabatic local spin density approximation~\cite{ALSDA}, fail to capture this ultranonlocal character, limiting their ability to describe spin-charge dynamics. To address this limitation, several studies~\cite{PhysRevB.62.4853,PhysRevB.68.045307,LFCforWeber} have proposed parameterizations based on insights gleaned from the random phase approximation (RPA) of current-current response functions. These response functions relate to the dynamic spin susceptibility through gauge symmetry. Notably, Reference~\cite{PhysRevB.68.045307} demonstrated that the SCD effect remains finite in the RPA of the current-current response function, even without taking into account XC corrections. This finding highlights the importance of the current-current response function as an alternative approach to studying the SCD effect. However, due to the lack of first-principles calculations of the spin XC kernel, these parameterizations rely on analytical ansatzes and approximations. While these approximations provide valuable insights, they may introduce systematic errors that are difficult to quantify. 

To obtain a comprehensive understanding of the XC kernel, it is imperative to compute the frequency-resolved spin susceptibility for the UEG problem using first-principles numerical methods. While the static spin susceptibility in the UEG has been extensively studied with quantum Monte Carlo (QMC) methods~\cite{LFCGFMC2D,LFCGFMC3D,ceperley1980VMC,2DGFMC,DMC1994spin,SENATORE2001333,umrigar1999quantum,PhysRevB.64.233110,PhysRevB.104.195142,PhysRevB.107.L201120}, the dynamic case remains far more challenging. Recent efforts have employed path-integral QMC to calculate the dynamic spin susceptibility in Matsubara frequency at high temperatures~\cite{DornheimGm}, but the accessible temperature range is insufficient to reveal the low-frequency structure of the spin XC kernel. This low-frequency behavior is crucial for understanding the SCD effect and developing reliable parameterizations for TDDFT calculations.

In this work, we develop a Variational Diagrammatic Monte Carlo (VDMC) method \cite{chen2019combined,haule2022single} for calculating the dynamic spin susceptibility of the 3D UEG in the Matsubara frequency representation. By reaching sufficiently low temperatures, we precisely calculate the dynamical structure of the spin XC kernel $K_{\rm XC}^-(q,i\omega_n)$ in both low and intermediate frequency regimes. Our numerical results across various electron densities confirm the theoretically predicted ultranonlocal singularity $A(i\omega_n)/q^2$ in the long-wavelength limit. For comparison, we calculate the charge channel XC kernel $K_{\rm XC}^+(q,i\omega_n)$, where this singularity is notably absent. 

Furthermore, by performing least-squares fitting of our $K_{\rm XC}^-(q,i\omega_n)$ data to an ansatz based on Eq.~\eqref{Eq:Low-freq singularity}, we extract the low-frequency components of $A(i\omega_n)$. The first-two order coefficients in the frequency expansion of $A(i\omega_n)$ directly determine the spin mass enhancement and spin diffusion relaxation time, respectively, as shown in Eq.~\eqref{eq:XCofsmsd}. Our calculations reveal a pronounced trend where the spin mass enhancement factor increases substantially as the electron density decreases, which are consistent with the previous results derived from the RPA methods but yield higher precisions \cite{PhysRevLett.93.106601}. Additionally, our results demonstrate vanishing inverse spin diffusion relaxation times, providing direct evidence for the suppression of spin diffusion processes at low temperatures as previous study has predicted \cite{PhysRevB.62.4853}.  Our study paves the way for the development of reliable $\emph{ab initio}$ methods for the spin dynamics in real materials.

The structure of our paper is as follows. Section~\ref{SecII} details the VDMC method for calculating the dynamic spin susceptibility in the 3D UEG. In section~\ref{SecIIIA}, we demonstrate the ultranonlocal behavior of the spin XC kernel. In section~\ref{SecIIIB}, we extract precise value of the spin mass enhancement factor and spin Coulomb drag relaxation time, discussing their dependence on electron density and temperature. Finally, Section~\ref{SecIV} summarizes our conclusions.

\section{Model \& Method}
\label{SecII}
\begin{figure}[b]
    \centering   \includegraphics[width=1.0\linewidth]{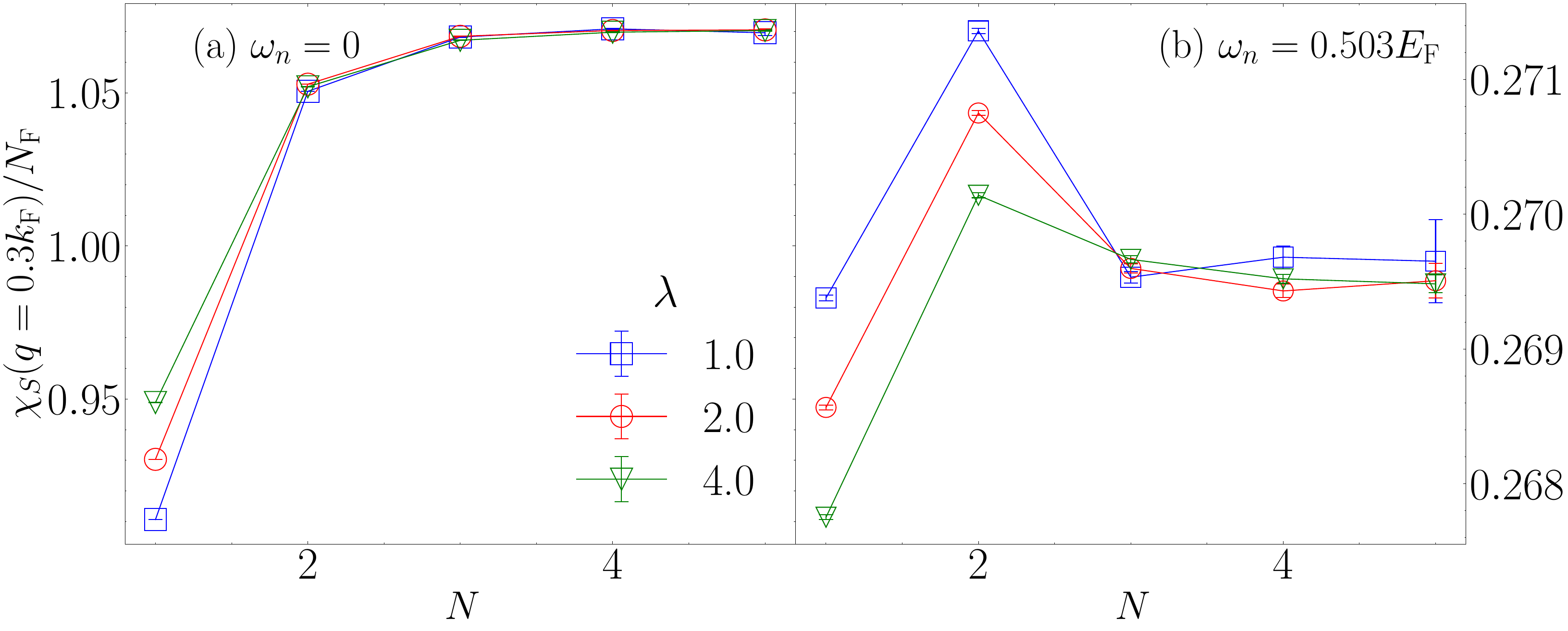}
    \caption{Spin susceptibility $\chi_S$ at $q=0.3k_{\rm F}$ versus truncation order $N$ for $\theta = 0.01$ and $r_s$ = 1. Panel~(a) shows the static case, and Panel~(b) shows the case with $\omega_n=0.503E_{\rm F}$. All $\lambda$ choices lead to the same extrapolated value, and the optimal $\lambda$ for the fastest convergence is about 2.0 for $\omega_n = 0$ and 4.0 for $\omega_n = 0.503E_{\rm F}$. Here $k_{\rm F}$ denotes the Fermi momentum, $E_{\rm F}$ denotes the Fermi energy and $N_{\rm F}$ denotes the density of state at the Fermi surface in the non-interacting system.}
    \label{fig:conv}
\end{figure}

We focus on an interacting electron system modeled as a UEG without any disorder. This system consists of electrons uniformly distributed within a homogeneous, positively charged background, interacting via the Coulomb potential. The system is conveniently described by two essential parameters: the density parameter, often referred to as the Wigner-Seitz radius $r_s=\bar{r}/a_B$, and the reduced temperature $\theta = T/T_{\rm F}$. Here, $\bar{r}$ denotes the average interparticle distance, $a_B$ the Bohr radius, and $T_{\rm F}$ is the Fermi temperature. Besides, there are some characteristic constants of the system, such as the Fermi momentum $k_{\rm F}$, the Fermi energy $E_{\rm F}$, and the density of state at the Fermi surface in the non-interacting system $N_{\rm F}$. The Hamiltonian governing the dynamics of this system is expressed as follows:

\begin{align}
    H = & \sum_{\vk\sigma}(\vk^2-\mu)\psi^\dagger_{\vk\sigma}\psi_{\vk\sigma} \notag  \\ 
    &+\frac{1}{2}\sum_{\vq\neq0\atop\vk\vk^\prime\sigma\sigma^\prime}\frac{8\pi}{\vq^2}\psi^\dagger_{\vk+\vq\sigma}\psi^\dagger_{\vk^\prime-\vq\sigma^\prime}\psi_{\vk^\prime\sigma^\prime}\psi_{\vk\sigma},
    \label{eq:UEG_Hamitonian}
\end{align}
where $\psi$,$\psi^\dagger$ are the annihilation and creation operators of a quasi-electron, $\mu$ is the chemical potential that is controlled by the parameter $r_s$,  and the Hamiltonian is formulated using Rydberg atomic units.

Addressing the many-body problem of the UEG Hamiltonian poses significant challenges due to the divergences arising from the bare Coulomb interaction in the diagrammatic expansion~\cite{VanHoucke2020}. To overcome this issue, we employ the Variational Diagrammatic Monte Carlo (VDMC) method~\cite{chen2019combined,haule2022single,diagMC1, diagMC2, kozik, diagMC4, boldDiagMC, rossi2018, van2012feynman,hou2024feynman} 
, an advanced field-theoretic approach that offers controlled accuracy. The VDMC method transforms the problem into an equivalent and more appropriate form for the expansion, taking the emergent low-energy physics as the lowest order of the model. This transformation significantly improves the convergence of the expansion with increasing perturbation order.

Within the VDMC framework, the system's action $S$ is decomposed into a reference action $S_0$ and a sequence of counterterms, serving as corrections. The Coulomb interaction inherent in the system is replaced by the Yukawa interaction $8\pi/(q^2+\lambda)$, with $\lambda$ serving as a variational screening parameter. This substitution allows for the representation of physical observables, such as electronic polarization, through a renormalized Feynman diagrammatic series~\cite{PhysRevB.93.161102,peskin2018introduction}
, expanding in powers of the Yukawa term. The introduction of the “polarization” counterterm $\lambda/8\pi$ effectively cancels out large contributions arising from particle-hole fluctuations, expediting the convergence of the diagrammatic series. The parameter $\lambda$ is subject to iterative optimization to enhance convergence~\cite{kleinert2009path}.

The computational framework of VDMC also incorporates chemical potential counterterms to preserve the electron density at each expansion order. Based on a self-consistent Hartree-Fock (HF) solution for the Green's function, the diagrammatic series is further simplified by omitting Fock-type self-energy insertions. We optimize the electron potential $v_{\vk}$ by inserting the GW-type self-energy, the Fock subdiagram, as the zeroth-order of the effective potential into the bare electron propagator. For higher orders of $v_{\vk}$, we add chemical-potential counterterms to fix the Fermi surface at each order, ensuring that the electron density remains unchanged order by order, in accordance with the Luttinger theorem.

High-order diagrams are efficiently evaluated through a Monte Carlo method employing importance sampling, with the sampling efficiency optimized using a computational graph representation of the diagrams~\cite{hou2024feynman,Baozong2021Ferminon}. The VDMC methodology has been successfully applied to explore various properties of the UEG, including the static and dynamic exchange-correlation kernel~\cite{chen2019combined, PhysRevB.104.195142, DynamicsofXC}, the effective mass~\cite{haule2022single}, and the behavior of the electron gas under extreme conditions~\cite{PhysRevB.106.L081126}. By optimizing the diagrammatic expansion, VDMC achieves reliable infinite-order results for any quantity without the need for a large truncation order $N$, significantly reducing computational costs while ensuring rapid and precise convergence for high-order calculations of physical observables. 

In our investigation, we employ VDMC to evaluate the imaginary-time spin-spin correlation function $\chi_S(q, \tau) = \left<\mathcal{T}\hat{s}_z(q, \tau)\hat{s}_z(0)\right>$ in the thermodynamic limit. Subsequent Fourier transformation yields the correlation function in Matsubara frequency space. We then compare the dynamical spin correlation function with theoretical predictions given by Eqs.~\eqref{DefKXC}~-~\eqref{eq:XCofsmsd} to probe the real-frequency dynamics of the spin susceptibility.

To validate the VDMC methodology, we calculate the spin susceptibility $\chi_{S}$ at $r_s = 1, \theta=0.01, q=0.3k_{{\rm F}}$ up to the fifth diagrammatic order. Figure~\ref{fig:conv} illustrates rapid numerical convergence of $\chi_{S}$ in the vicinity of the optimal $\lambda$ for two characteristic Matsubara frequencies $\omega_n = 0$ and $\omega_n = 0.503E_{{\rm F}}$. This confirms VDMC's capability for high-precision, high-order calculations.

\begin{figure}[ht]
  \centering    
  \includegraphics[width=0.75\linewidth]{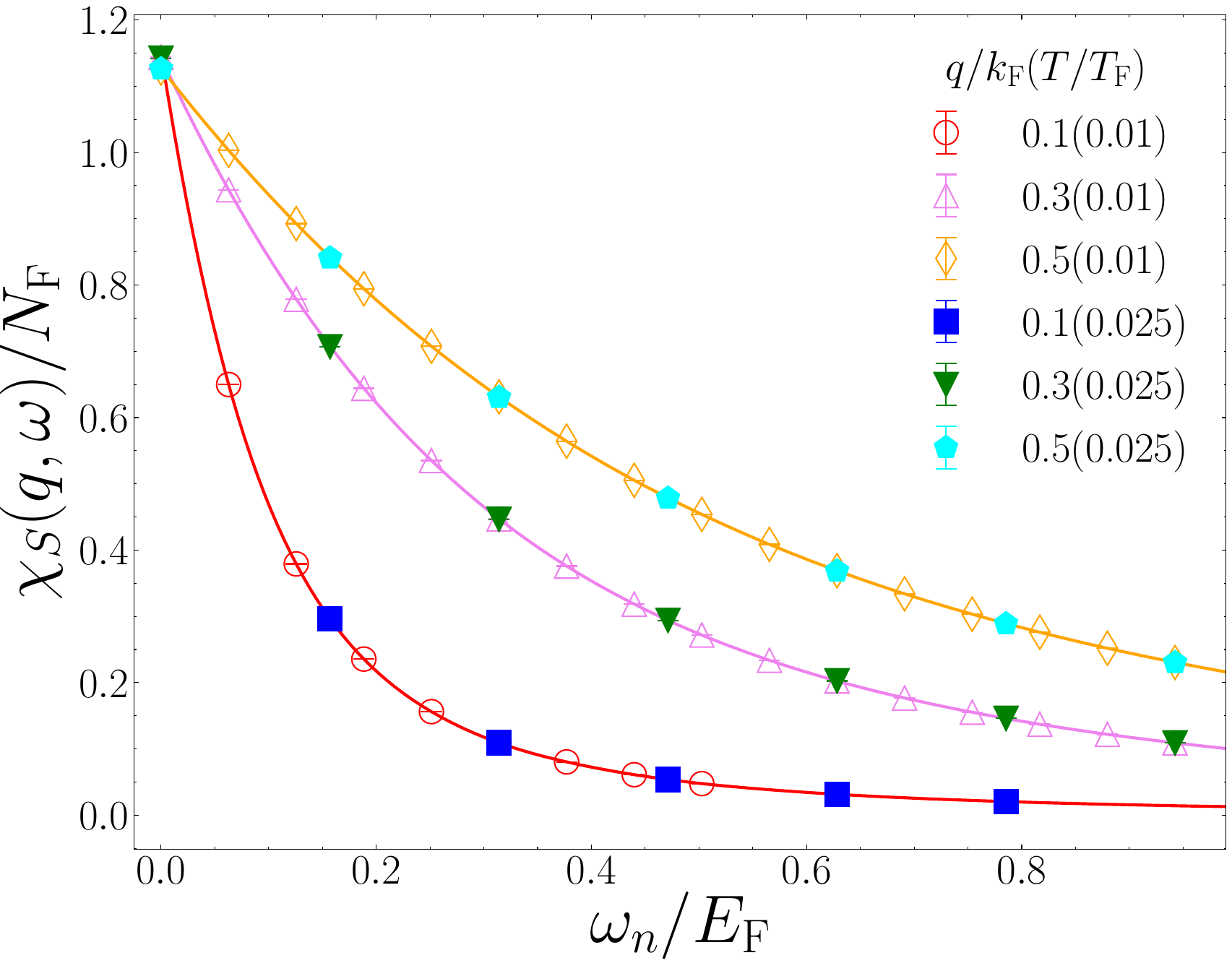} 
    \caption{Dynamic spin susceptibility $\chi_S$ as the Matsubara frequency $\omega_n$ increases for different temperatures ($\theta = 0.025, 0.01$) with varying $q$ of the UEG system ($r_s = 1.0$), indicating that the temperatures used in our simulations are effectively zero.}
    \label{fig:Temp-comp}
\end{figure}

\begin{figure}[t] 
    \centering    \includegraphics[width=1.02\linewidth]{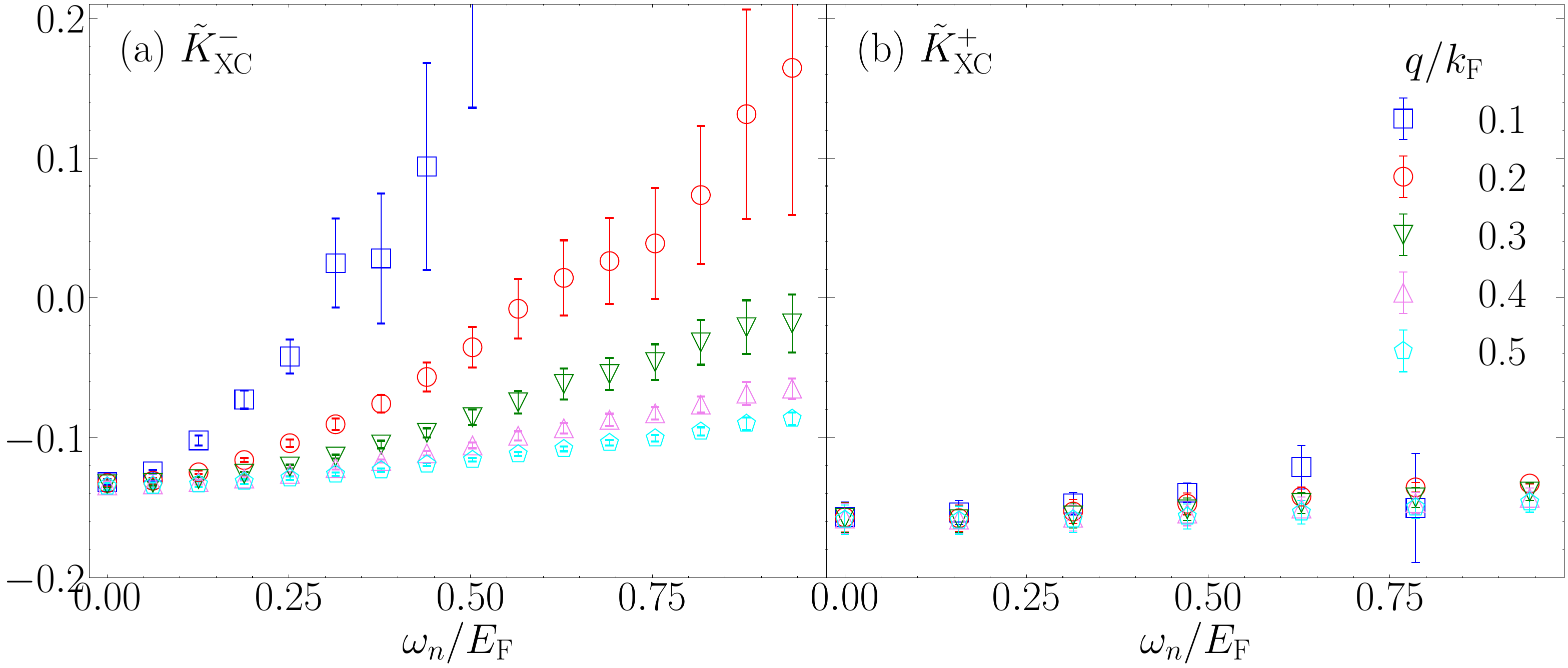} 
    \caption{(a) Spin exchange-correlation (XC) kernel with ($r_s =1,\theta = 0.01$). They show that the dynamic $K^-_{\rm XC}$ exhibits a universal divergence in the small-$q$ limit. (b) Charge XC kernel as $\omega$ increases with various $q$ with ($r_s =1,\theta=0.025$), showing that $K^+_{\rm XC}$ converges in the small-$q$ limit.  Here $\Tilde{K}_{\rm XC}^\pm := K_{\rm XC}^\pm N_{\rm F}$ is the reduced XC kernel.}
    \label{fig:charge-response}
\end{figure}

\section{Results}
We perform extensive VDMC simulations for 3D UEG systems with $r_s = 0.5, 1.0, 2.0$ at temperatures $\theta=0.025, 0.01$, measuring the spin susceptibility $\chi_S$ for momenta $q$ within $0.5k_{\rm F}$ and Matsubara frequencies $\omega_n \leq E_{\rm F}$. Figure~\ref{fig:Temp-comp} shows the collapse of $\chi_S(i\omega_n,\theta=0.025)$ and $\chi_S(i\omega_n,\theta=0.01)$ onto the same curve for a given $q$, indicating that these temperatures are sufficiently low to converge to the zero-temperature limit. Consequently, we focus on simulations at $\theta = 0.01 $ to obtain a denser frequency grid for detailed study of the spin response characteristics.

\subsection{Ultranonlocality in the Spin Channel}
\label{SecIIIA} 
A key manifestation of many-body effects in the spin response of the 3D UEG caused by the spin Coulomb drag is the ultranonlocal behavior of the dynamic XC kernel $K_{\rm XC}^-$. This ultranonlocality, often referred to as the `ultranonlocality problem' in time-dependent spin-density-functional theory \cite{PhysRevLett.90.066402}, is characterized by a divergence of $K_{\rm XC}^-$ in the small-$q$ limit at finite frequencies.

To investigate this ultranonlocal behavior, we compute $K_{\rm XC}^-$ through the dynamic spin susceptibility $\chi_S$ and the Lindhard function $\chi_0$ using Eq.~\eqref{DefKXC}.
Figure \ref{fig:charge-response}(a) shows the dimensionless $K_{\rm XC}^-$ as a function of $\omega_n$ for different momenta at $r_s = 1.0$. Apart from the static case ($\omega_n$ = 0), $K_{\rm XC}^-$ increases rapidly as $q$ approaches zero for a given frequency, clearly demonstrating the ultranonlocal behavior.

\begin{figure*}[t]
    \centering
    \includegraphics[width=1.0\linewidth]{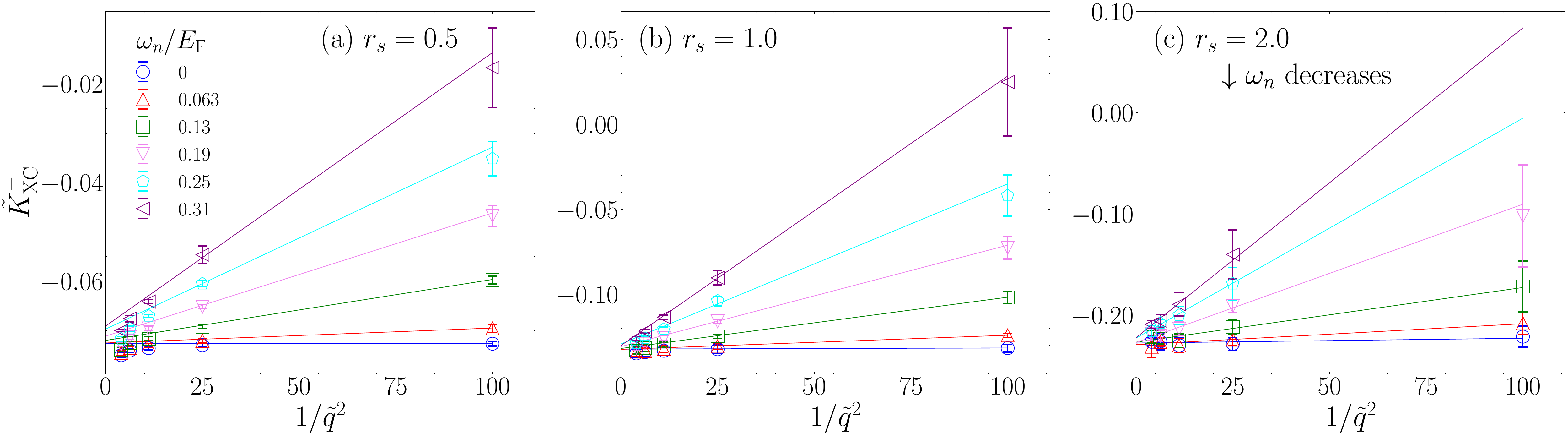}
    \caption{The reduced XC kernel $\Tilde{K}_{\rm XC}^-$ excluding the $\mathcal{O}(\redq^2)$ term versus $1/\redq^2$ ($\redq = q/k_{\rm F}$) with various frequency $\omega_n$  for (a) $r_s = 0.5$, (b) $r_s = 1.0$ and (c) $r_s = 2.0$. The slope of the solid lines $a$ is derived by fitting Eq.~\eqref{fit_Kxc} for each frequency. It implies that there is a $1/q^2$-divergence of the spin XC kernel, which demonstrate the ultranonlocal behaviors in the spin response.}
    \label{fig:omegafix}
\end{figure*}

To highlight the uniqueness of this behavior in the spin response, we compare it with the charge XC kernel $K_{\rm XC}^+$, defined as
\begin{equation}
  K_{\rm XC}^+ = {\chi_0}^{-1}(\bold{q},\omega) - \chi^{-1}_{nn}(\bold{q},\omega),  
  \label{XC-charge-definition}
\end{equation}
where $\chi_{nn}(q,\tau)= \left<\mathcal{T}\hat{n}(q, \tau)\hat{n}(0)\right>$ is the density-density response function. Figure~\ref{fig:charge-response}(b) shows that $K_{\rm XC}^+$ saturates to a constant in the small-$q$ region, in stark contrast to the ultranonlocal behavior of $ K_{\rm XC}^- $. This comparison reveals distinct behaviors in the interactions between electrons of different spins, with the ultranonlocality being a unique feature of the spin response. Furthermore, As one can see later in Fig.~\ref{fig:omegafix}, the ultranonlocal behavior of $ K_{\rm XC}^- $ is consistently observed at different electron densities ($ r_s = 0.5 $, $1.0$, and $ 2.0 $), substantiating its universality in the 3D UEG.

Importantly, our analysis of the singular behaviors of the XC kernel, as expressed in Eq.~\eqref{eq:XCofsmsd}, can be carried out in the Matsubara frequency representation ${\omega_n}$ at the effective zero temperature. This is made possible by the application of the Wick rotation, a technique that performs an analytical continuation from real frequencies to imaginary frequencies via the transformation $\omega+i0^+ \to i\omega_n$. The Wick rotation is a powerful tool in quantum field theory and many-body physics, allowing for the calculation of real-time quantities using imaginary-time formalisms \cite{altland2010condensed}.

The applicability of the Wick rotation in our case is justified by the fact that the XC kernel is analytic in the upper half of the complex frequency plane \cite{2005Quantum}. This analyticity property ensures that the imaginary-frequency representation contains the same information as the real-frequency one. Moreover, the Matsubara formalism is particularly advantageous for numerical calculations, as it avoids the singularities and branch cuts that may appear in the real-frequency domain \cite{bruus2004many}.

Applying the Wick rotation to Eq.~\eqref{eq:XCofsmsd}, we obtain:
\begin{equation}
A(\omega_n) = \frac{m}{n\tau_{sd}}\omega_n+\left(\frac{m_s}{m}-1\right)\frac{m}{n}\omega_n^2+\mathcal{O}\left(\left(\frac{\omega_n}{E_{\rm F}}\right)^3\right),
\label{eq:sm_sdmf}
\end{equation}
where the first two terms of interest are exactly real. These terms capture the leading contributions to the ultranonlocal behavior at low frequencies and relate the ultranonlocality to two fundamental properties of spin transport: the spin Coulomb drag effect and the spin mass enhancement.

To investigate the structure of the spin XC kernel, we perform a least-square fit of our numerical data to the analytical ansatz:
\begin{equation}
K_{{\rm XC}}^-(q,i\omega_n) \xrightarrow{q \to 0, i\omega_n \ll E_{\rm F}} \frac{A(i\omega_n)}{q^2}+B+\mathcal{O}(q^2, i\omega_n),
\end{equation}
where $B$ is a constant. This ansatz captures the leading terms in the small-$q$ limit of the XC kernel at finite but small Matsubara frequencies. To facilitate the fitting procedure, we introduce the dimensionless XC kernel $\Tilde{K}_{{\rm XC}}^- := K_{{\rm XC}}^-N_{{\rm F}}$ and the dimensionless momentum $\Tilde{q}:=q/k_{{\rm F}}$. We then perform the fit using the following equation:
\begin{equation}
\Tilde{K}_{{\rm XC}}^-(\Tilde{q};\omega_n) = \frac{a(\omega_n)}{\redq^2} +b(\omega_n),
\label{fit_Kxc}
\end{equation}
where $a(\omega_n)$, $b(\omega_n)$ are frequency-dependent fitting parameters.  Given that our data falls within the small-$q$ range ($0.1 \leq \Tilde{q} \leq 0.5$), we omit terms of $q^2$ and higher orders to ensure a stable fit.

The fitting is carried out for each Matsubara frequency $\omega_n$ within the range $0 \leq \omega_n \leq 0.314 E_{{\rm F}}$ and for three different electron densities corresponding to $r_s = 0.5$, $1$, and $2$. The obtained fitting parameters are reported in Table~\ref{Tab:Fit_1}.

\begin{table}[h]
\begin{tabular}{l|l l l}
\hline\hline
$r_s$ & $\omega/E_{\rm F}$ & \quad \quad $a$          & \quad \quad $b$                 \\ \hline
&    0.000	 & 0.000\,016(5)	 & -0.074\,0(2) \\ 
&    0.063	 & 0.000\,049(9)	 & -0.073\,9(2) \\ 
0.5&    0.126	 & 0.000\,145(9)	 & -0.073\,5(2) \\ 
&    0.188	 & 0.000\,304(17)	 & -0.073\,3(3) \\ 
&    0.251	 & 0.000\,45(3)	 & -0.072\,5(4) \\ 
&    0.314	 & 0.000\,73(5)	 & -0.072\,7(3) \\ 
      \hline
   &    0.000	 & 0.000\,02(3)	 & -0.133\,9(15) \\ 
&    0.063	 & 0.000\,096(19)	 & -0.133\,7(12) \\ 
1.0&    0.126	 & 0.000\,33(4)	 & -0.133\,4(10) \\ 
&    0.188	 & 0.000\,65(6)	 & -0.133\,0(11) \\ 
&    0.251	 & 0.001\,0(1)	 & -0.132\,3(11) \\ 
&    0.314	 & 0.001\,69(15)	 & -0.132\,5(13) \\ \hline
&    0.000	 & 0.000\,05(11)	 & -0.228(3) \\ 
&    0.063	 & 0.000\,21(12)	 & -0.229(3) \\ 
2.0&    0.126	 & 0.000\,5(2)	 & -0.227(4) \\ 
&    0.188	 & 0.001\,4(3)	 & -0.227(5) \\ 
&    0.251	 & 0.002\,2(7)	 & -0.223(6) \\ 
&    0.314	 & 0.003(1)	 & -0.222(6) \\ 
\hline
\end{tabular}
\caption{Fitting result of $K_{\rm XC}^-$ for various frequency via Eq.~\eqref{fit_Kxc} for various $r_s$.}
\label{Tab:Fit_1}
\end{table}

To validate the consistency of our numerical results with the analytical ansatz, we examine the behavior of the modified XC kernel, $\Tilde{K}_{{\rm XC}}^- - c(\omega_n)\redq^2$, which excludes the regular $\mathcal{O}(q^2)$ term. As shown in Fig.~\ref{fig:omegafix}, this modified XC kernel exhibits a clear linear dependence on $1/q^2$ for various $\omega_n$ values, confirming the presence of the $1/q^2$-dominant term in the ultranonlocal behavior of the 3D UEG, as predicted by the theory.

The slopes of the linear fits in Fig.~\ref{fig:omegafix}, represented by the solid lines of different colors, correspond to the fitting parameter $a(\omega_n)$ for each Matsubara frequency. Notably, the slopes increase monotonically with increasing frequency, indicating that $a(\omega_n)$ is a monotonically increasing function of $\omega_n$ within the low-frequency domain, without any divergence.

\begin{figure*}[t]
    \centering
    \includegraphics[width=1.0\linewidth]{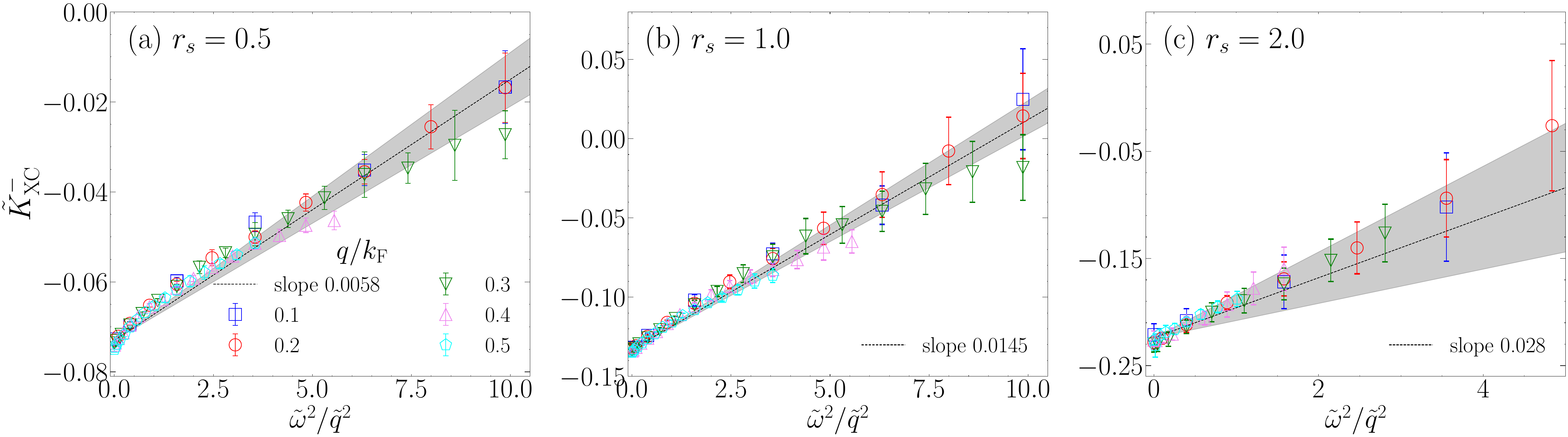}
    \caption{The reduced XC kernel versus $\Tilde{\omega}^2/\Tilde{q}^2$ for various $r_s$: (a) 0.5, (b) 1.0, (c) 2.0. The slope of the dashed lines represent the parameter $a_2$ and the shaded area represents its error, which implies the spin mass renormalization term is dominant at low temperature while the spin diffusion is vanishing.}
    \label{fig:RegXC}
\end{figure*}

The comprehensive analysis of our numerical data through the least-square fitting procedure demonstrates the consistency of the ultranonlocal behavior of the spin XC kernel in the 3D UEG with the analytical predictions. The frequency-dependent fitting parameters obtained from this analysis will serve as a foundation for the investigation of the spin diffusion and the spin mass enhancement in the following subsection. 

\subsection{Spin Mass Enhancement and Spin Diffusion Phenomenon}
\label{SecIIIB}
The frequency-dependent fitting parameter $a(\omega_n)$, obtained from the analysis of the ultranonlocal behavior of the spin XC kernel, provides valuable insights into two fundamental properties of spin transport in the 3D UEG: the spin mass enhancement and the spin diffusion phenomenon. These properties are encapsulated in the low-frequency expansion of $a(\omega_n)$, as indicated by Eq.~\eqref{eq:sm_sdmf}.

To quantitatively investigate these properties, we perform a least-square fit of $a(\omega_n)$ using a second-order polynomial ansatz in terms of the reduced Matsubara frequency $\redomega := \omega/E_{\rm F}$:
\begin{equation}
    a(\omega) = a_0 + a_1 \redomega + a_2 \redomega^2+O(\redomega^3).
    \label{eq:fit_a}
\end{equation}
The constant term $a_0$ is set to zero, as the XC kernel $K_{\rm XC}^-$ converges for different momenta $q$ in the static limit ($\omega = 0$), as shown in Fig.~\ref{fig:charge-response}. The fitting coefficients $a_1$ and $a_2$ are determined for three different electron densities, corresponding to $r_s = 0.5$, $1.0$, and $2.0$, and are reported in Table~\ref{Fit_result}.

By comparing the low-frequency expansion of $a(\omega_n)$ in Eq.~\eqref{eq:fit_a} with the analytical expression in Eq.~\eqref{eq:sm_sdmf}, we can extract the spin diffusion relaxation time $\tau_{sd}$ arising from the spin Coulomb drag effect and the dimensionless spin mass enhancement factor $m_s/m$:
\begin{align}
    \tau_{sd} &= \frac{mE_{\rm F}N_{\rm F}}{na_1k_{\rm F}^2}, \\
    \frac{m_s}{m} &= \frac{nk_{\rm F}^2a_2}{mN_{\rm F}E_{\rm F}^2} + 1.
\end{align}
The values of $\tau_{sd}$ and $m_s/m$, computed using the fitting coefficients, are also listed in Table~\ref{Fit_result} for each $r_s$.

\begin{table}[h]
\begin{tabular}{l|ll|cc} \hline
$r_s$ & $a_1(10^{-4})$       & \quad $a_2$    & $E_{\rm F}^{-1}/\tau_{sd} (10^{-4})$ &\quad $m_s/m$  \\  \hline
0.5   & $4.3(12)$&0.005\,8(6) & $6(2)$  &1.007\,7(8)  \\ 
1.0   & $7(3)$ & 0.014\,5(19) & $9(4)$  &1.019\,3(3) \\
2.0   & 10(20) & 0.028(12) & - & 1.04(2)\\\hline
\end{tabular}
\caption{Fitting coefficients $a_1$ and $a_2$ extracted from the low-frequency expansion of $a(\omega_n)$ for different $r_s$ values, along with the derived spin Coulomb drag relaxation time $\tau_{sd}$ (normalized by the inverse Fermi Energy $E_{\rm F}^{-1}$) and the spin mass enhancement factor $m_s/m$.}
\label{Fit_result}
\end{table}

The spin mass enhancement factor $m_s/m$ quantifies the renormalization of the electron mass due to many-body effects in the spin channel. Our numerical analysis yields highly precise values for the spin mass enhancement factor $m_s/m$: $1.007\,7(8)$ for $r_s = 0.5$, $1.019\,3(3)$ for $r_s = 1.0$, and $1.04(2)$ for $r_s = 2.0$. These results corroborate previous theoretical predictions of $m_s/m = 1.02$ and $1.06$ for $r_s = 1.0$ and $2.0$ respectively \cite{PhysRevLett.93.106601}, which were obtained from the calculation of the dynamic spin XC kernel through a decoupling approximation of the four-point density function combined with diffusion Monte Carlo-calculated static local field factors \cite{PhysRevB.64.233110,XC3dspin}. Our calculations achieve substantially higher precision, enabling more rigorous tests of theoretical models against experimental measurements. The systematic increase in spin mass enhancement with $r_s$ indicates strengthening many-body effects as the electron system becomes more strongly correlated. 

The relaxation time of the spin diffusion $\tau_{sd}$ characterizes the decay of spin currents due to electron-electron interactions. In the low-temperature limit, $\tau_{sd}$ is expected to diverge in the 3D UEG, as the phase space for electron-electron scattering vanishes \cite{PhysRevB.62.4853,2005Quantum}. Our numerical results confirm this behavior, as the ratio of the inverse Fermi energy to $\tau_{sd}$, is found to be of the order of $10^{-4}$ for $r_s = 0.5$ and $1.0$. For $r_s = 2.0$, the uncertainty in $a_1$ exceeds its value, making a reliable determination of $\tau_{sd}$ challenging. Nevertheless, the overall trend suggests that $1/\tau_{sd}$ approaches zero as the temperature tends to zero, consistent with the theoretical expectation of vanishing spin diffusion in the 3D UEG at zero temperature.

The dominance of the spin mass enhancement over the dissipation of the spin current caused by the spin Coulomb drag effect in the low-temperature limit is further corroborated by the plot of the reduced XC kernel $\Tilde{K}_{\rm XC}^-$ as a function of $\redomega^2/\redq^2$ for different $r_s$ values, as shown in Fig.~\ref{fig:RegXC}. The linearity of the plots, with slopes given by the fitting coefficient $a_2$, demonstrates that the spin mass renormalization term is the leading contribution to the ultranonlocal behavior of the XC kernel in this regime.

Through systematic analysis of the frequency-dependent parameter $a(\omega_n)$, our VDMC calculations quantify the spin mass enhancement and spin diffusion suppression in the three-dimensional uniform electron gas at low temperatures. The precise determination of these quantities demonstrates the capability of VDMC to resolve subtle many-body effects in spin transport, providing quantitative benchmarks for future theoretical and computational studies.

\section{Discussions}
\label{SecIV}
In this work, we have employed the VDMC approach to investigate the spin susceptibility of the 3D UEG at low temperatures. Our study has focused on the spin Coulomb drag effect through the ultranonlocality behaviors of  the spin-resolved XC kernel and its connection to fundamental properties of spin transport, namely the spin mass enhancement and the spin diffusion phenomenon.

Through extensive VDMC simulations, we have computed the dynamic spin susceptibility and the associated XC kernel for various electron densities, momenta, and Matsubara frequencies. Our results clearly demonstrate the presence of a $1/q^2$ divergence in the spin XC kernel at finite frequencies, confirming the spin Coulomb drag effect predicted by theoretical studies. Remarkably, this singularity is absent in the charge channel, highlighting the unique nature of many-body effects in the spin response.

By fitting our numerical data to an analytical ansatz for the XC kernel, we have extracted the frequency-dependent coefficient of the $1/q^2$ term, which encodes information about the spin mass enhancement and the dissipation of the spin current. Our analysis yields precise values for the spin mass enhancement factor, aligning closely with previous theoretical predictions and significantly improving upon their accuracy.  We observe an increasing trend in spin mass enhancement with decreasing electron density, highlighting the growing importance of many-body effects in the strongly correlated regime. Furthermore, our study presents numerical evidence for the suppression of spin diffusion in the 3D UEG at low temperatures. The extracted relaxation times of the spin current are found to be several orders of magnitude larger than the inverse Fermi energy, suggesting that spin currents can persist for extended durations in this system. This finding is consistent with the theoretical expectation of vanishing spin diffusion in the zero-temperature limit~\cite{2005Quantum}.

Looking ahead, the insights gained from this work may guide the development of more accurate density functional approximations for spin-dependent phenomena, with potential applications in spintronics and quantum technologies. Future extensions of our methodology to more complex systems, such as multicomponent electron gases, cold atomic gases, and realistic materials, hold promise for further advancing the field.

 \section*{Acknowledgements}
K. C. was supported by the National Key Research and Development Program of China, Grant No. 2024YFA1408604, and the National Natural Science Foundation of China under Grants No. 12047503 and No. 12447103. Z.L, P.H., and Y.D. were supported by the National Natural Science Foundation of China (under Grant No. 12275263), the Innovation Program for Quantum Science and Technology (under grant No. 2021ZD0301900), the Natural Science Foundation of Fujian Province of China (under Grant No. 2023J02032).

\bibliography{ref}
\end{document}